\documentclass[aps,prl,floats,twocolumn,showpacs,superscriptaddress]{revtex4-1}
\usepackage{graphicx,epsfig}
\usepackage{times}
\usepackage{graphics,dcolumn,bm,fleqn,epic,eepic,float}
\usepackage{amssymb,amsmath,multirow,rotate,color}
\usepackage{color}
\begin{document}
\title{Networks of motifs from sequences of symbols}

\author{Roberta Sinatra,$^{1,2,*}$ Daniele Condorelli,$^{2,3}$ and Vito Latora}
\affiliation{Dipartimento di Fisica ed Astronomia, Universit\`a di
  Catania, and INFN, Via S. Sofia 64, 95123 Catania, Italy}
  \affiliation{Laboratorio sui Sistemi Complessi, Scuola Superiore di
  Catania, Via San Nullo 5/i, 95123 Catania, Italy
  \\$^{3}$Dipartimento di Scienze Chimiche, Sezione di Biochimica e
  Biologia Molecolare, Universit\`a di Catania, Viale A. Doria 6,
  95125 Catania, Italy}

\email[Corresponding author: ]{roberta.sinatra@ct.infn.it}

\date{\today}

\begin{abstract}
  We introduce a method to convert an ensemble of sequences of symbols
  into a weighted directed network whose nodes are motifs, while the
  directed links and their weights are defined from statistically
  significant co-occurences of two motifs in the same sequence. The
  analysis of communities of networks of motifs 
  is shown to be able to correlate sequences with functions in
  the human proteome database, to detect hot topics from online social
  dialogs, to characterize trajectories of dynamical systems, and
  might find other useful applications to process large amount of data
  in various fields.
\end{abstract}

\maketitle

There are many examples in biology, in linguistics and in the theory
of dynamical systems, where information resides and has to be
extracted from corpora of raw data consisting in sequences of symbols.
For instance, a written text in English or in another language is a
collection of sentences, each sentence being a sequence of the letters
from a given alphabet. Not all sequences of letters are possible,
since the sentences are organized on a lexicon of a certain number of
words. In addition to this, different words are used together in a
structured and conventional way \cite{ramon, motter}.  Similarly, in
biology, DNA nucleotides or aminoacid sequence data can be seen as
corpora of strings \cite{searls,trifonov86,peng,scafetta}. For
example, it is well known that proteomes are far from being a random
assembly of peptides, since clustering of aminoacids \cite{rosato} and
strong correlations among proteomic segments \cite{bussemaker} have
been clearly demonstrated. These results give meaning to the metaphor
of protein sequences regarded as texts written in a still unknown
language \cite{searls,solan}. Sequences of symbols can also be found
in time series generated by dynamical systems. In fact, a trajectory
in the phase space can be transformed into sequence of symbols, by the
so-called ``symbolic dynamic'' approach \cite{beck}. The basic idea is
to partition phase space into a finite number of regions, each of
which is labelled with a different symbol. In this way, each initial
condition gives rise to a sequence of symbols representing the initial
cell, the cell occupied at the first iterate, the cell occupied at the
second iterate, and so forth.

In all the examples mentioned above, the main challenge is to decipher
the message contained in the corpora of data sequences, and to infer
the underlying rules that govern their production.
In order to do this, one needs: 
{\em i)} to detect the fundamental units carrying information,
like words do in language, and {\em ii)} to study their combination
syntax in the ensemble of sequences. In fact, information in its
general meaning is located not only at the level of strings, but also
in their correlation patterns \cite{lacasa, tadic}. 
In this Letter, we introduce a method to transform a generic corpus of
strings, such as written texts, protein sequence data, sheet music, a collection
of dance movement sequences \cite{bradley10}, into a network representing the significant and fundamental
units of the original message together with their relationships.  The
method relies on a statistical procedure to detect patterns carrying
relevant information, and works as follows. We first construct a
dictionary of the recurrent strings of $k$ letters, called
$k$-motifs. Recurrent strings play, in this more general context, the
same role as words in written or spoken languages. We then construct a
$k$-motif network, a graph in which each node is one entry of the
dictionary, and a directed arc between two nodes is drawn when the
ordered co-occurence of the two motifs is statistically significant in
the dataset analyzed. We will show how the analysis
  of topological properties of networks of $k$-motifs, such as the
  detection of community structures \cite{bocca,fortunato}, allows to
  extract important information encoded in the original data. In
  particular, we will consider the application of the method to
  datasets in three different domains, namely, biological sequences of
  proteins, messages from online social networks, and sequences of
  symbols generated by the trajectories of a dynamical system.

Let us consider an ensemble $\cal S$ of $S$ sequences of symbols.
Each sequence $s$ ($s=1,2,\ldots,S$) is a string of letters from an
alphabet $\cal A$ of $A$ letters, ${\cal A} \equiv \{ \sigma_1
,\sigma_{2},...,\sigma_A \}$. In general, the strings can have
different lengths. We indicate by $l_s$ the length of sequence $s$,
and by $L=\sum_{s=1}^S l_s$ the total length of the ensemble. An
example is provided by proteomes. A proteome is a
collection of $S \approx 10^4$ proteins of a species. Each protein is 
a sequence of length $l_s$, ranging from $10^2$ to $10^3$, made of
symbols from an alphabet $\cal A$ with $A=20$ letters, ${\cal A}
\equiv \{ \sigma_1 ,\sigma_{2},...,\sigma_{20} \}$, where each
$\sigma$ labels one of the aminoacids a protein can be made
of. We define as $k${\em -string} a segment of $k$ contiguous
letters $x_1x_2 \ldots x_k$, where $x_i \in {\cal A} ~ \forall i$. The
number of all possible $k$-strings is $A^k$, while from the ensemble
of sequences $\cal S$ we can select only $L-S\cdot(k-1)$ overlapping
$k$-strings, so that some of the possible $k$-strings do not occurr,
some of them occur once, others more than once, either in the same or in different sequences of symbols. We define as:
\begin{equation}
  p^{obs}(x_1x_2 \cdots x_k)  = \frac    { c(x_1 x_2 \cdots x_k)                       }
  {\sum_{(x_1, x_2, \cdots, x_k) \in{\cal A}^k}  c(x_1 x_2 \cdots x_k) }
\label{p-obs}
\end{equation}
the {\em observed probability} of a string $x_1 x_2 \ldots x_k$. This
probability is obtained by counting the total number of times, $c(x_1
x_2 \ldots x_k)$, the string actually occurs in the sequences of the
ensemble.  To assess for the statistical significance of the string,
the probability in Eq.~\ref{p-obs} has to be compared with the {\it
  expected probability} $p^{exp}(x_1x_2 \cdots x_k)$ of the string 
occurrence. The latter can be evaluated under different
assumptions. In fact, the joint probability $p(x_1 x_2 \cdots x_k)$
can be written as:
\begin{equation}
 p(x_1 x_2 \cdots x_k)=p(x_1 x_2\cdots x_{k-1})p(x_k|x_1x_2\cdots x_{k-1}),
\nonumber
\end{equation}
and different approximations for the conditional probabilities
$p(x_k|x_1x_2\cdots x_{k-1})$ lead to different values of the expected
probability $p^{exp}(x_1x_2\cdots x_k)$. Namely, if we assume that the
occurrence of a letter does not depend on any of the previous letters,
i.e. $p(x_k|x_1x_2\cdots x_{k-1})=p(x_k)$, the expected probability is
simply given by the product of the relative frequencies of the
string's component letters: $p^{exp}(x_1x_2\cdots
x_k)=p^{obs}(x_1)\cdots p^{obs}(x_k)$ \cite{ferraro,giansanti}. By
using instead a first order Markov approximation,
i.e. $p(x_k|x_1x_2\cdots x_{k-1})=p(x_k|x_{k-1})$, the expected
probability can be expressed in the form: $p^{exp}(x_1x_2 \cdots x_k)
= p^{obs}(x_1) p^{obs}(x_2|x_1) \cdots p^{obs}(x_k|x_{k-1})$, where
$p^{obs}(x_j|x_i)$ is extracted from the countings as:
$p^{obs}(x_j|x_i) = c(x_i x_j) / \sum_{x_j} c(x_i x_j) = p^{obs}(x_i
x_j)/p^{obs}(x_i)$.  This latter assumption is based on the fact that
there is a minimal amount of memory in the sequence: a symbol of the
sequence is correlated to the previous one only.  Here, we go beyond
the approximation of Markov chains of order 1, by retaining as much
memory as possible \cite{trifonov86}.  We assume:
\begin{eqnarray}
\nonumber
p^{exp}(x_1 x_2 \cdots x_k) = p^{obs}(x_1 x_2 \cdots x_{k-1}) \cdot \\
\cdot p^{obs}(x_k|x_2 \cdots x_{k-1})
\end{eqnarray}
where the conditional probabilities can be evaluated from countings as:  
\begin{eqnarray}
p^{obs}(x_k|x_2 \cdots x_{k-1}) = \frac{c(x_2 x_3 \cdots x_k)} {\sum_{x_k} c(x_2 x_3 \cdots x_k)}
\end{eqnarray}
or can be expressed in terms of the observed probability for shorter
sequences as:
\begin{eqnarray}
  p^{obs}(x_k|x_2 \cdots x_{k-1}) =  \frac{ p^{obs}(x_2 \cdots x_k)}{p^{obs}(x_2 \cdots x_{k-1})}
\end{eqnarray}
By using the latter expression, we can finally write the expected probabilities in a more compact form:
\begin{eqnarray}
\label{beyond2}
p^{exp}(x_1)             &=& p^{obs}(x_1)
\nonumber\\ 
p^{exp}(x_1 x_2)         &=& p^{obs}(x_1 x_2) 
\nonumber\\ 
p^{exp}(x_1 x_2 x_3)     
			&=& p^{obs}(x_1 x_2) \frac{p^{obs}(x_2 x_3)}{p^{obs}(x_2)}
\nonumber
\\      ...                &=& ....
\\
\nonumber
p^{exp}(x_1 x_2 \cdots x_k) 
%
&=& p^{obs}(x_1 \cdots x_{k-1})\cdot \\
\nonumber
&\;& \cdot \frac{p^{obs}(x_2 \cdots x_k)} {p^{obs}(x_2 \cdots x_{k-1})}\\
\nonumber
\end{eqnarray}
This way, the expected probability of a given
$k$-string is evaluated based on observations for strings of up to
$(k-1)$ symbols. Therefore, by predicting the probability of
appearance with a high order Markov model, our method allows to
highlight the true $k$-body correlations subtracting from them the
effects due to $(k-1)$ and lower order correlations. Based 
on observed and expected probabilities, a test
of statistical significance, for instance a $Z$-score, is then performed
for each $k$-string. We define {\em $k$-motifs} or
\emph{recurrent $k$-strings}, the statistically-relevant strings whose
observed and expected number of occurrences are such as to validate
the statistical test adopted, and we indicate as ${\cal Z }_k$ the 
dictionary composed by all the selected $k$-motifs \cite{urialon}.

Once we have constructed a lexicon of fundamental units, the next goal 
is to represent in a graph the way they are combined together. 
Recurrent $k$-strings can be distributed differently along the 
sequences: they can appear in single sequence or in more than one
sequence, alone or in clusters. To extract the non trivial patterns of
correlated appearance of $k$-motifs, we need to evaluate the
probability for the random co-occurrence of two motifs, when these are
uncorrelated.  We estimate first the expected probability that motif
$X$ is followed by motif $Y$ within a generic sequence of the ensemble
$\cal {S}$, then we sum over all the sequences of $\cal {S}$. We
denote as $p(X)$ and $p(Y)$
the probabilities of finding the two motifs in $\cal {S}$. In
sequence $s$, motif $X$ can occupy positions ranging from the first to
the $(l_s-2k)$th site, where $l_s$ is the length of $s$, and $k$ is the length of the motif. We have assumed that the two motifs cannot overlap. For each fixed
position $i$ of $X$ on $s$, with $i=1,...,(l_s-2k)$, there are
$(l_s-2k+1-i)$ possibilities for $Y$ to appear in the sequence. Hence, 
the number of expected co-occurences of $X$ and $Y$ within $s$ is given
by: $\sum_{i=1}^{l_s-2k} (l_s-2k+1-i) p(X) p(Y)$. In order to obtain
the expected number of co-occurrences, we have to sum over all the sequence in the ensemble $\cal
{S}$. We finally
get:
\begin{equation}
\begin{split}
 N^{exp}(Y|X) = p(X) p(Y)\sum_{s=1}^S \sum_{i=1}^{l_s-2k} (l_s-2k+1-i)  = \\
\negthickspace \negthickspace \negthickspace = \frac{1}{2} p(X)p(Y) \sum_{s=1}^S (l_s-2k+1)(l_s-2k+2) 
\end{split}
\end{equation}
%
For each value of $k$, we are now able to construct the {\em $k$-motif
  network} of the ensemble $\cal S$, i.e. a directed network whose nodes are motifs in the
dictionary ${\cal Z }_k$, and an arc point from node $X$ to node $Y$
if the number of times $Y$ follows $X$ in the ensemble of sequences is
statistically significant. Furthermore, a weight can be associated to
the arc from $X$ to $Y$, based on the extent to which the
co-occurrence of the two motifs deviates from expectation.
 
%
This approach is able to represent the correlation
  patterns encrypted in the ensemble of sequences into a single
  object, the $k$-motif network. Then, graph theory allows to extract
  information from the structural properties of the network, and
  to retrieve the main message encoded in the original sequences. In
  particular, it is interesting to study the components of the
  $k$-motif network or, if the graph is connected, its community
  structures, i.e. those groups of nodes tightly connected among
  themselves and weakly linked to the rest of the graph
  \cite{fortunato}.

\begin{figure}[!t]
\begin{center}
\epsfig{file=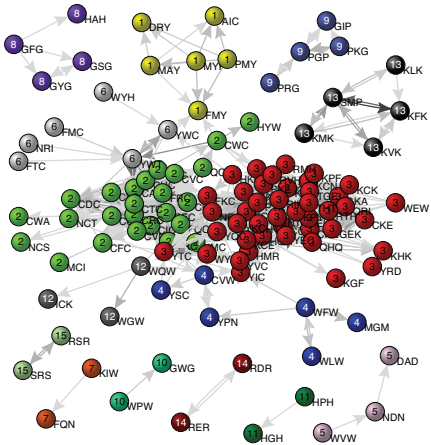, width=0.33\textwidth,angle=0,clip=1}
\end{center}
\caption{\label{network} The 3-motifs network of the human proteome. 
  Nodes belonging to the same community are labeled by the same number
  and share the same colour. Most of the communities can be associated
  to a functional domain as described in table I in \cite{supplementary}.}
 \end{figure}
 In the following, we will consider the application
   of the method to three different datasets, belonging to three
   contexts as diverse as biology, social dialogs and dynamical
   systems. We will show how the community analysis of the related
   $k$-motif networks enables to extract functional domains in
   proteomes, social cascades and hot topics in Twitter, and the
   increase of chaoticity in deterministic maps.

 In the biological context, many methods based on
   strings deviating from expectancy in genome
   \cite{trifonov86,caselle} or in a proteome \cite{nicodeme} have
   been already used to make functional deductions. Although they
   provide insight on many biological mechanisms \cite{giansanti},
   this approach turns out to be not sufficient for a complete and
   exhaustive interpretation of the genomic and proteomic message. A
   fundamental key to its comprehension is in fact hidden in the
   correlations among recurrent patterns of strings, which are
   perfectly represented at a global scale in terms of $k$-motif
   networks. Various features of these correlations translate into
   structural properties of $k$-motif networks. In Fig.~\ref{network}
   we illustrate, as an example, the 3-motif graph derived from the
   ensemble of human proteins (see \cite{supplementary} for details
   about the dataset). We have detected 15 different
   communities in the graph, labeled in the figure with different colours and
   numbers. By means of a research in biological databases, we can show
   that linked couples of motifs belonging to the same
   community all co-occur in the same kind of protein domains and that one can associate 9 of these 15 communities just
   to one domain (see table I in
   \cite{supplementary}).
   These results are outstanding compared to
   the current methods to extract functional protein domains, all
   based on multi-alignment of sequences, and cannot obtained 
   if one uses a lower order Markov model, meaning that it is fundamental to 
   take into account both short- and long-range correlations (for more details on the
   $k$-motif networks in proteomes, see
   \cite{supplementary}).

\begin{table}[!t]
  \caption{\label{table_tweets} 
    The first ten most significant links between motifs, belonging to 7 different communities
    in the Twitter dataset \cite{supplementary}. Each community corresponds 
    to a specific tweet or expression that generated a topic cascade. }
\begin{center}
\begin{tabular}{ c | c | c | p{3cm} |p{2.4cm} }
\hline
\hline

 \hspace{1pt} motif\hspace{1pt} & \hspace{1pt} motif \hspace{1pt} & $\frac{p^{obs}}{p^{exp}}$  & \multicolumn{1}{c|}{Expression or Tweet} & \multicolumn{1}{c}{Topic}\\

 1 & 2 &  &   & \\
 \hline
 9cle & gg27 & 955.3 & \multirow{3}{3cm}{ \emph{GUARDIAN ICM POLL Cameron 35\% Brown 29\% Clegg 27\%} } &\multirow{7}{2.4cm} {poll results from various websites, journals, tv channels, etc}\\
5bro	&	wn29	&	894.8	&			&	\\
& & & & \\
\cline{1-4}
son4	&	4cle	&	924.3	&	\multirow{4}{3cm}{\emph{Brown wins on 44\%, Clegg is second on 42\%, Cameron 13\% None of them 1\%}}	&		\\
don4	&	2cam	&	881.7	&			&	\\
& & & & \\
& & &  & \\
\hline
lapo	&	mete	&	892.3 & www.slapometer.com &	\multirow{2}{2.4cm}{A funny website on the election} \\
& & & & \\
\hline
swed	&	nesd	&	864.7	&	 \multirow{5}{3cm}{\emph{hey Dave, Gordon and Nick : how about a 4th debate on Channel 4 this wednesday night without the rules?!}} & \multirow{3}{2.4cm} {Proposal for a 4th debate among leaders, made by a journalist on his twitter page}\\
nesd	&	ayni	&	826.1	&	&			\\
 & & & 	&\\
& & & & \\
& & & & \\
\hline
jami	&	ncoh	&	842.0	&	Benjamin Cohen	&		\multirow{2}{2.4cm} {Journalist of Channel 4 News}	\\
minc	&	ohen	&	764.9	&		&			\\
\hline
isob	&	eymu	&	831.4	&	\#disobeymurdoch	&		hashtag	\\
 \hline
 \hline
   \end{tabular}
   \end{center}
\end{table}

Important information from $k$-motif networks can
  also be retrieved from datasets of social dialogs and microblogging
  websites, like Twitter. Although in these cases, in principle, a
  dictionary is a-priori known, not all terms used in the Internet
  language are always listed in the dictionary \cite{mitrovic}:
  abbreviations, ``leet language'' words, names of websites or 
  of public personages, are just some examples. Moreover, some
  expressions or combinations of terms appear more frequently in some
  periods or contexts due to the interest in some hot topics. We have
  found that communities of $k$-motif networks derived from
  microblogging sequences in Twitter during the UK Election in April 2010
  are able to detect exactly those
  hot topics which generate information cascades
  \cite{twitter_cascades}, as shown in Fig.1 and Table II of 
  \cite{supplementary}. In Table \ref{table_tweets} we report the links with the highest significance
  together with the tweet associated to their community. Each tweet was the origin 
  of a cascade and can be associated with a specific topic or event
  discussed during the election campaign (see \cite{supplementary} for details).

\begin{figure}[! tb]
\begin{center}
\epsfig{file=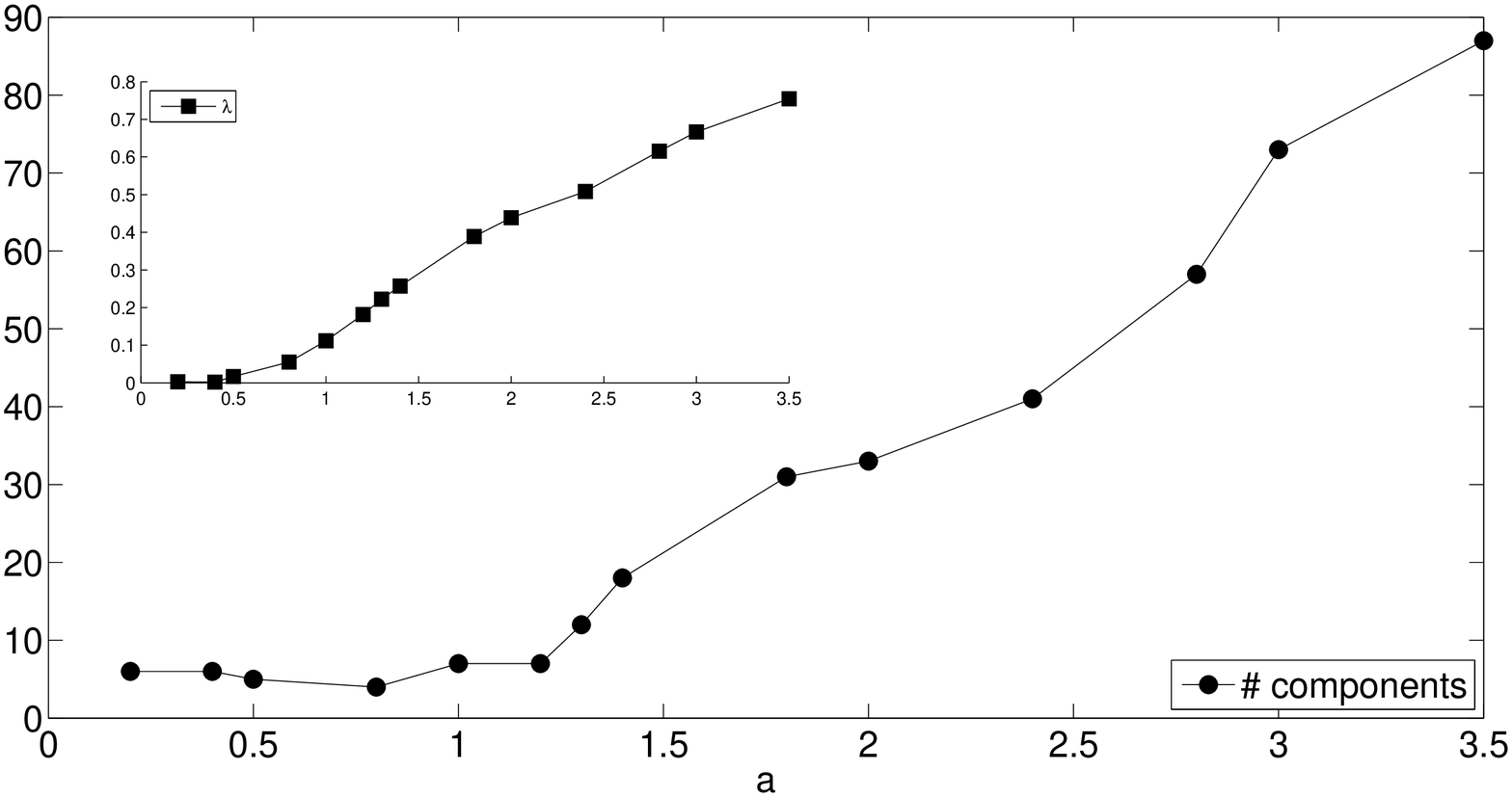, width=0.4\textwidth,angle=0,clip=1}
\end{center}
\caption{\label{lyapunov_standard_map} Standard map: number of components in
  the $3$-motifs networks (main figure), and the Lyapunov exponent (inset), as a
  function of the non-linearity parameter $a$. }
 \end{figure}

Finally, $k$-motif networks carry important
  information on sequences of symbols generated from trajectories of
  dynamical systems by the so-called ``symbolic dynamic'' approach
  \cite{beck}. One is able, for instance, to distinguish ensembles of
  sequences generated by deterministic maps from those generated by
  stochastic processes, by looking at the number of components and
  communities in the $k$-motif network. In fact, the method, when
  applied to sequences generated by deterministic equations that are
  increasingly non-linear, still finds short motifs, while the same
  does not occur for ensembles of random sequences.  Furthermore, we
  have found that the higher is the non-linearity in a conservative
  deterministic dynamical system, the more disconnected is the
  corresponding $k$-motif network.  In
  Fig.~\ref{lyapunov_standard_map}, we show an example of this
  behaviour for a well-known two-dimensional area-preserving
  deterministic map, the standard map \cite{chirikov}. Each point in
  Fig.~\ref{lyapunov_standard_map} represents the number of components
  in the $3$-motif network obtained from an ensemble of trajectories
  produced for a specific value of the non-linearity parameter $a$.
  We observe that the number of components increases with $a$, and
  this behavior is similar to that of the positive Lyapunov exponent
  of the map, shown in the inset (see also \cite{supplementary}). 

\medskip

Summing up, in this Letter we have introduced a general method to
construct networks out of any symbolic sequential data. The method is
based on two different steps: first it extracts in a ``natural'' way
motifs, i.e. those recurrent short strings which play the same role
words do in language; then it represents correlations of motifs within
sequences as a network. Important information from the original data
are embedded in such a network and can be easily retrieved as shown
with different applications (a biological system, a social
  dialog and a dynamical system). With respect to
previous linguistic methods, our approach does not need the a priori
knowledge of a given dictionary,  and also allows to
  compare different ensembles, corresponding, for example, to
  different values of control parameters in dynamical systems.  All
this makes the method very general and opens up a wide range of
applications from the study of written text, to the analysis of sheet
music or sequences of dance movements. Moreover, the method 
does not use parameters on the position of
motifs in order to correlate them, since co-occurrences are computed
within sequences, which represent natural interruptions of a corpora
of data (proteins in a proteome, posts in a blog, different initial
conditions in a symbolic dynamics, etc.).

\medskip
\begin{acknowledgments}
  \small We thank A. Giansanti and V. Rosato for stimulating
  discussions on the biological applications of the method,
  S. Scellato for providing us with the Twitter dataset, and M. De
  Domenico for his interesting comments on applications to dynamical
  systems.  This work was partially supported by the Italian To61 INFN
  project.

\end{acknowledgments}

\newpage

\Large{Supplementary material to ``Networks of motifs from sequences of symbols''}

\bigskip

\normalsize
The properties of $k$-motif networks can reveal important
characteristics of the message encrypted in the original data, as the
analysis of topological quantities (clustering coefficient, average
path length and degree distributions) has helped to understand various
linguistic features in networks of words co-occuring in sentences
\cite{ramon, caldeira}, and also to model how language has evolved in
networks of conceptually-related words \cite{motter}.  We present here
details and further results on the application of the method described in the
main article to three different datasets: proteomic
sequences, short text messages acquired from \emph{Twitter}, the
well-known social network and microblogging platform, and ensembles of
sequences derived from dynamical trajectories of the standard map 
by means of a symbolic dynamics approach.

\section{Biological sequences}
Methods to study over- or under-representation of particular motifs in a complete genome
\cite{trifonov86,caselle} or in a proteome \cite{nicodeme}, have
already been proposed, and the results have been used to make
functional deductions.  Although the information contained in strings
deviating from expectancy is useful for the analysis of many
biological mechanisms \cite{giansanti}, it turns out to be not
sufficient for a complete and exhaustive interpretation of the genomic
and proteomic message. A fundamental key to its comprehension is in
fact hidden in the correlations among recurrent patterns of
strings. The spatial structure of proteins provides an example: when a
protein folds, segments distant on the sequence come to be close to
each others in the space. This can happen because two (or more)
segments need to physically interact in order to perform the
biological function the protein is supposed to go through. Such a
mechanism translates into a statistical correlation between short
motifs of aminoacids, which is well captured by an analysis in terms
of $k$-motif networks. 

\begin{table}[!hb]
  \caption{\label{table_community} 
  List of communities in the 3-motif network of the human
    proteome. Community labels as in Fig.~1 of the main text,
    number of nodes, total internal weight, 
    associated domain, and the domain specificity are reported.}
\begin{center}
\begin{tabular}{ c | c c c c}
\hline
\hline
 \hspace{1pt} \hspace{1pt} & \hspace{1pt} \# nodes \hspace{1pt}& \hspace{1pt} Internal  \hspace{1pt} & \hspace{1pt} Domain \hspace{1pt} & Domain \\

 \hspace{1pt}  \hspace{1pt} & \hspace{1pt} \hspace{1pt}& \hspace{1pt} weight \hspace{1pt} &  \hspace{1pt}  \hspace{1pt} & recognition \\
 \hline
 1	&	6	&	83,30\%		&	Olfactory 	& 171/175 \\
 	&		&			&	 receptor	&  \\
 \hline
 2	&	25	&	74,91\%		&	---		&\\
 \hline
 3	&	43	&	94,13\%		&	Zinc Finger	& 1345/1364 \\
 \hline
 4	&	6	&	55,42\%		&	G-protein and	& 9/11 \\
 	&		&			&	CUB-Sushi 	&  \\
 \hline
 5	&	3	&	100\%		&	Cadherin	&  330/347\\
 \hline
 6	&	4	&	100\%		&	Lipoproteins	& 16/19\\
 \hline
 7	&	2	&	100\%		&	Homeobox &  65/84 \\
 \hline
 8	&	4	&	100\%		&	---	&  \\
 \hline
 9	&	4	&	100\%		&	Collagen 	& 271/482 \\
 \hline
 10	&	2	&	100\%		&	Serine 		& 22/51 \\
 	&		&			&	protease	&  \\
 \hline
 11	&	2	&	100\%		&	---		&  \\
 \hline
 12	&	3	&	60,30\%		&	C-type 		& 3/4\\
 	&		&			&	proteins	& \\
 \hline
 13	&	5	&	100\%		&	---	& \\
 \hline
 14	&	2	&	100\%		&	---	& \\
 \hline
 15	&	2	&	100\%		&	---	& \\
 \hline
 \hline
   \end{tabular}
\end{center}
\end{table}

 \begin{figure*}[!ht]
\begin{center}
\epsfig{file=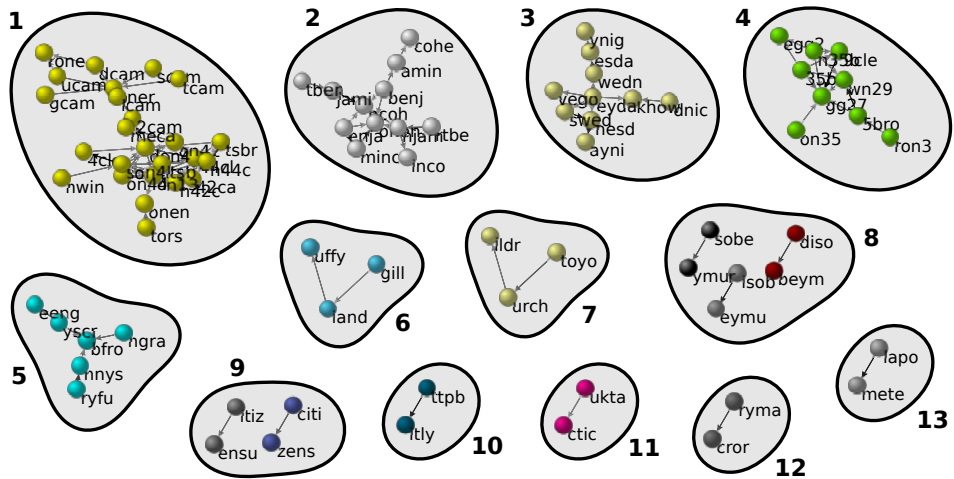, width=0.7\textwidth,angle=0,clip=1}
\end{center}
\caption{\label{motifs_election} Components of the $4$-motifs network of the twitter dataset. Each component and its associated topic are described in table \ref{table_tweets}.}
 \end{figure*}
 
\subsection*{Human proteome}
In our application, we have considered the ensemble of sequences
relative to the human proteome \cite{CCDS}.  It consists of 34180
aminoacidic sequences of variable size, with an average length of 481
letters.  For this dataset, we have computed the probabilities
$p^{obs}$ and $p^{exp}$ for each of the $20^3=8000$ possible strings
of three aminoacids, and we have selected as $3$-motifs the strings
satisfying $\frac{p^{obs}}{p^{exp}}>\left\langle
  \frac{p^{obs}}{p^{exp}} \right\rangle +2\sigma$, hence creating the
dictionary $Z_3$ \cite{average}. The entries of the dictionary are the nodes
of the 3-motif network.  The node $X$ is then linked to $Y$ with a
directed arc if the number of times that motif $Y$ follows motif $X$
within the same protein is statistically significant, according to the
relation: $\frac{p^{obs}(Y|X)}{p^{exp}(Y|X)}> \left\langle
  \frac{p^{obs}(Y|X)}{p^{exp}(Y|X)} \right\rangle + 2 \sigma$. The
statistical significance $\frac{p^{obs}(Y|X)}{p^{exp}(Y|X)}$ is also
the weight of the arc. In this way we obtain the 3-motif graph of 199
nodes and 1302 directed links, shown in Fig. 1 of the main
article. The graph has 86 isolated nodes (not displayed in Figure),
while the remaining 113 nodes are organized into 10 weak components.
The largest component of the graph contains 5 clusters, detected by
means of the MCl algorithm \cite{vandongen}. Therefore, 15 different
communities are present in the graph. In Table \ref{table_community}
we report, for each community, the number of nodes and its total
internal weight, defined as the sum of the weights of links between
nodes of the communities normalized by the sum of the weights of links
incident in nodes of the community.  By submitting a query to the
Prosite database \cite{prosite} we have obtained, for each couple of
connected motifs belonging to the same community, the list of all
proteins, classified by domain, where the two motifs co-occur.  The
results show that linked couples of motifs belonging to the same
community, all co-occur in the same kind of domains. In addition to
this, one can associate 9 of these 15 communities just to one protein
domain, since the majority of co-occurrences emerge in proteins
matching a well-defined function. In Table \ref{table_community} we
report, when possible, the association to a single protein domain,
together with the ratio between the number of times the couple of
motifs with the highest weight occurred in that specific domain, and
the total number of co--occurrences in the database.

Analogous results were also found for the 4-motif graph \cite{article_preparation}, 
while it is not possible to derive the same kind of information by using 
lower order Markov models  to construct dictionaries. For example, the 3-motif 
network constructed with a dictionary based on a lower order approximation rather than
on a 2-bodies Markov chain, exhibits a community structure with just four communities, 
none of which could be identified with a functional protein domain.

\section{Social networks and microblogging}
By means of $k$-motif networks, information can also be retrieved from
datasets of social dialogs and microblogging websites. Although in
these cases, in principle, a dictionary is a-priori known, not all
terms used in the Internet language are always listed in a dictionary:
abbreviations, puns, leet language words \cite{leet}, names of websites or names
of public figures, are just some examples. Moreover, some expressions
or combinations of terms appear more frequently in some periods or
contexts due to the interest to some hot topics. In addition to this,
the method of $k$-motif networks turns to be very useful in all those
contexts where it is necessary to process and compact information from
large amount of symbolic data. This is the case of Internet, where the
amount of text data provided by blogs, dialogs in social networks,
forums, etc. is growing and growing.

In the following, we provide details on how network of motifs are able
to deduce information about hot topics and cascades
\cite{twitter_cascades, flicker_cascades} in a dataset extracted from
Twitter, a well-know platform for social networking and microblogging.

\subsection*{Twitter}
\emph{Twitter} \cite{twitter} is a social networking and microblogging
service which allows users to send short messages known as
\emph{tweets}. Tweets are composed only of text, with a strict limit
of 140 characters: they are displayed on the author's profile page and
delivered to the author’s subscribers, who are also known as
``followers''.  The dataset we have analyzed is a collection of 28143
tweets, crawled on two days, from the 23rd to 24th April 2010, and
selected through the Twitter Streaming API \cite{API} if they
contained the string \emph{\#leadersdebate}. The choice of such a
keyword, called in Twitter also \emph{hashtag}, was aimed to select
all those tweets concerning electoral campaign in UK, where general
election to elect the members of the House of Commons would have taken
place two weeks later.  We have analyzed the dataset removing all
blank spaces between words and all symbols that where not numbers or
letters (punctuation, symbols like \$, @, *, etc.) and not
distinguishing between lower- and upper-case
letters. 
From these sequences, dictionaries of motifs $ {\cal Z}_3$ and ${\cal
  Z}_4$ have been extracted, selecting respectively the 10\% and 1\%
of most significant strings of $3$ and $4$ letters. As described in
the main text, we have constructed networks whose nodes represent the
entries of a dictionary, and an arc is drawn from the node
representing string X to the node standing for string Y, if
$p^{obs}(Y|X)/p^{exp}(Y|X)$ is greater than a certain threshold. In
Fig.~\ref{motifs_election}, we show the $4$-motifs network when the
threshold is set equal to 400 (isolated nodes not reported). Such a
high threshold is chosen to have a small network that can be easily
visualized and studied. More information can be obtained by setting
the threshold to lower values or analyzing networks made up of motifs
of different length $k$. Searching in the original dataset the
connected motifs, it is possible to associate each component to a
particular tweet which generated a cascade or with a specific
expression, related to a specific hot topic discussed by users of the
microblogging platform.  For all components of
Fig.~\ref{motifs_election}, we report in Table~\ref{table_tweets} the
tweet or expression associated and its meaning. For example, component
1 and 4 can be associated to two exit polls disclosed on those days by
two different journals, or component 6 to the name ``Gillian Duffy'',
a 65-years old pensioner involved in a political scandal with British
PM Gordon Brown during the election tour (Brown's remarks of her as a
``bigoted woman'' were accidentally recorded and broadcast).

\section{Symbolic dynamics}
Symbolic dynamics is a general method to transform trajectories of
dynamical systems into sequences of symbols. The distinct feature in
symbolic dynamics is that time is measured in discrete intervals. So
at each time interval the system is in a particular state. Each state
is associated with a symbol and the evolution of the system is then
described by a sequence of symbols. The method turns to be very useful
in all those cases where system states and time are inherently
discrete. In case the time scale of the system or its states are not
discrete, one has to set a coarse-grained description of the system.
Different initial conditions usually generate different trajectories
in the phase space, which map onto different sequences of symbols. A
large number of initial conditions produces an ensemble of sequences
whose analysis can be addressed with the method based on networks of
motifs, as described in the main article.

In the following, we will describe the application of the method to the
standard map, and we will show how the related networks of
motifs shape according to its chaotic behavior.
 
 \begin{table}[!ht]\footnotesize
  \caption{\label{table_tweets} In relation to Fig. \ref{motifs_election}, we report  the number of nodes, links, the tweet or the expression containing the motifs and the topic associated to each of the 13 communities }
\begin{center}
\begin{tabular}{ c | c | c | p{3cm} |p{2.5cm} }
\hline
\hline

 \hspace{1pt} Comm.\hspace{1pt} & \hspace{1pt} Nodes \hspace{1pt} & Links  & \multicolumn{1}{c|}{Expression or Tweet} & \multicolumn{1}{c}{Topic}\\
 \hline
1 &	25	&	33	&\emph{Brown wins on 44\%, Clegg is second on 42\%, Cameron 13\% None of them 1\%}	& poll results from various websites, journals, tv channels, etc	\\
\hline
2 &	12	&	14 	& Benjamin Cohen 			 &	 Journalist of Channel 4 News	\cite{cohen} \\ 
\hline
3 &	10	&	11	&	\emph{hey Dave, Gordon and Nick : how about a 4th debate on Channel 4 this wednesday night without the rules?!} & Proposal for a 4th debate among leaders, made by a journalist on his Twitter page\\
\hline
4 & 	9 	& 	13 &  \emph{GUARDIAN ICM POLL Cameron 35\% Brown 29\% Clegg 27\%}  & poll results from various websites, journals, tv channels, etc\\
\hline
5	&	6	 &	5	 &	\emph{Very funny screengrab from the LeadersDebate} & About a funny picture of the leaders debate on BBC \cite{screengrab} \\
\hline
6	&	3	&	2	&	Gillian Duffy	&	Woman branded a 'bigot' by Gordon Brown in general election campaign \cite{duffy}		\\
\hline
7	&	6	&	5	&	\emph{Cameron: I believe that if you've inherited hard all your life you should pass it on to your children}	&	Electoral campaign from David Cameron	\\
\hline
8	&	6 	&	3 	&	\#disobeymurdoch		& 		Twitter hashtag	\\
\hline
9	&	4	 &	2 &	\#citizensuk & Twitter hashtag \\
\hline
10	&	2	 &	1 &	http:// ... .ly & Format of shortened weblinks in twitter\\
\hline
11	&	2	 &	1 &	Tactical voting 			& \emph{Strategy that when a voter misrepresents his or her sincere preferences in order to gain a more favorable outcome \cite{tactical}}\\
\hline
12	&	2	 &	1 &	Henry Macrory  & Head of press for the Conservatives, owner of a twitter account \\
\hline
13	&	2	 &	1 &	www.slapometer.com   							 & A funny website on the election \\
 \hline
 \hline
   \end{tabular}
   \end{center}
 \end{table}
 
\subsection*{Standard Map}
The standard map, also known as Chirikov map, is a bidimensional area-preserving
chaotic map. It maps a square with side $2\pi$ onto itself
\cite{chirikov}. It is described by the equations:
\begin{eqnarray}
\left\lbrace 
\begin{array}{l}
x_{t+1}=p_t+a\sin{x_t} \; \mod{2\pi}\\
p_{t+1}=p_t+x_{t+1} \; \; \; \;\mod{2\pi}
\end{array}
\right. 
\label{eq:standard_map}
\end{eqnarray}

where $t$ represents time iteration and $a$ is a parameter assuming
real values. The map is increasingly chaotic as $a$ increases (see
inset of Fig. 2 in the main article to see a plot of the Lyapunov
exponent as a function of the parameter $a$).  For $a = 0$, the map is
linear and only periodic and quasiperiodic orbits are allowed. When
evolution of trajectories are plotted in the phase space (the
\emph{xp} plane), periodic orbits appear as closed curves, and
quasiperiodic orbits as necklaces of closed curves whose centers lie
in another larger closed curve. Which type of orbit is observed
depends on the map's initial conditions. When the nonlinearity of the
map increases, for appropriate initial conditions it is possible to
observe chaotic dynamics.


In order to obtain sequences from the standard map (\ref{eq:standard_map}) by means of the symbolic dynamic approach \cite{aleksev}, one needs to make a coarse graining of the phase space, defining a discrete and finite number of possible states the trajectory can occupy. This way it is possible to associate a symbol to each of the possible states and derive a sequence according to the trajectory originating from an initial condition. We have coarse-grained the phase space into 25 ($5\times 5$) squares of equal size and we have derived for different values of the parameter $a$, $10^4$ sequences of $10^3$ symbols. In other words, this means to follow for $10^3$ time steps the trajectories originating from $10^4$ different initial conditions.

The idea is that closed orbits or quasi periodic-ones correspond to correlations between motifs and therefore in links of the graph of motifs. When the map becomes more and more chaotic, closed orbits disappear and, correspondingly, the networks break in many components. In the extreme limit of map highly chaotic ($a>3$), the network of motifs are completly disconnected, with all nodes isolated. Nevertheless, this scenario is different from the one generated by stochastic sequences, since in this case motifs would not be detected, while this still happens in the chaotic map, although only for small values of $k$. This result is well depicted in Fig. 3 of the main article, where the number of components of the $3$-motif graphs is plotted as a function of the value $a$ of the map generating the ensemble. This curve is shown to have the same behavior of the Lyapunov exponent, as reported in the inset of the same figure.

 
%

\medskip
%


\begin{thebibliography}{99}

\bibitem{ramon} R. Ferrer i Cancho, R.V. Sol\'e and R. K\"ohler, \emph{Phys. Rev. E} {\bf 69}, 051915(R) (2004); R. Ferrer i Cancho and R.V. Sol\'e, \emph{Proc. R. Soc. Lond. B} {\bf 268}, 2261 (2001).

\bibitem{motter} A. Motter, A.P.S. De Moura, Y.C. Lai, and P. Dasgupta, \emph{Phys. Rev. E}, {\bf 65}, 065102 (2002); E.G. Altmann, J. B. Pierrehumbert and A.E. Motter, \emph{PLoS One} {\bf 4}:e7678 (2009).

\bibitem{searls} D. B. Searls, \emph{Nature} {\bf 420}, 211 (2002).

\bibitem{trifonov86} V. Brendel, J.S. Beckmann, and E.N. Trifonov, \emph{Journal of Biomolecular Structure \& Dynamics} \textbf{4}, 011 (1986).

\bibitem{peng} C.-K. Peng, S. V. Buldyrev, A. L. Goldberger, S. Havlin, F. Sciortino, M. Simons, H. E. Stanley, \emph{Nature} {\bf 356}, 168 (1992).

\bibitem{scafetta} N. Scafetta, V. Latora, and P. Grigolini, \emph{PRE} {\bf 66}, 031906 (2002). 

\bibitem{rosato} V. Rosato, N. Pucello, and G. Giuliano, \emph{Trends in Genetics} {\bf 18}, 278 (2002). 

\bibitem{bussemaker} H.J. Bussemaker, H. Li and E.D. Siggia, \emph{Proc. Natl. Acad. Sci. USA} {\bf 97}, 10096 (2000).

\bibitem{solan} Z. Solan, D. Horn, E. Ruppin and S. Edelman, \emph{Proc. Natl. Acad. Sci. USA} {\bf 102}, 11629 (2005).

\bibitem{beck} C. Beck and F. Schl\"ogl, Thermodynamics of chaotic systems (Cambridge University Press, Cambridge, 1993).

\bibitem{lacasa} L. Lacasa, B. Luque, F. Ballesteros, J. Luque, and J. C. Nunos, \emph{Proc. Natl. Acad. Sci. USA} {\bf 105}, 4972 (2008).

\bibitem{tadic} J. Zivkovic, M. Mitrovic and B. Tadic, \emph{Studies in Computational Intelligence}, {\bf 207}, Complex Networks, Eds. S. Fortunato et. al., 23-34, Springer, (2009).

\bibitem{bradley10} E. Bradley, D. Capps, J. Luftig, and J. Stuart, \emph{Open AI Journal} {\bf 4}, 1-19 (2010).

\bibitem{bocca} S. Boccaletti, V. Latora, Y. Moreno, M. Chavez, and D.-U. Hwang, \emph{Phys. Rep.} {\bf 424}, 175 (2006).

\bibitem{fortunato} S. Fortunato, \emph{Phys. Rep.} {\bf 486}, 75 (2010).

\bibitem{ferraro} L. Ferraro, A. Giansanti, G. Giuliano, and V. Rosato, arXiv:q-bio/0410011v2.

\bibitem{giansanti} A. Giansanti, M. Bocchieri, V. Rosato, and S. Musumeci, \emph{Parasitol. Res.} {\bf 101}, 639 (2007).

\bibitem{urialon} The term \emph{motif} is chosen in analogy with the concept of network \emph{motifs}, i.e. recurrent patterns of nodes and links in a graph. U. Alon, \emph{Nature Reviews Genetics} {\bf8}, 450 (2007).

\bibitem{caselle} M. Caselle, F. Di Cunto, and P. Provero, \emph{BMC Bioinformatics} {\bf 3}, 7 (2002); D. Cor\`a, F. Di Cunto, P. Provero, L. Silengo, and M. Caselle, \emph{BMC Bioinformatics} {\bf 5}, 57 (2004). 

\bibitem{nicodeme} P. Nicod\`eme, T. Doerks, and M. Vingron, \emph{Bioinformatics} {\bf 18}: S161, Suppl.2 (2002).

\bibitem{vandongen} A. J. Enright, S. Van Dongen, and C. A. Ouzounis, \emph{Nucleic Acids Research} {\bf 30}:1575 (2002).

\bibitem{supplementary} See supplementary material at the end of the paper for details and other results about the application of the method in the three datasets.

\bibitem{twitter_cascades} K. Lerman and R. Ghosh, in Proc. of ICWSM (2010).

\bibitem{mitrovic} M. Mitrovic and B. Tadic, \emph{Eur. Phys. J. B} {\bf 73}: 293 (2010).

\bibitem{chirikov} B.V. Chirikov, Phys. Rep. {\bf 52}:263 (1979).

\end{thebibliography}

\begin{thebibliography}{99}

\bibitem[S1]{ramon} R. Ferrer i Cancho and R.V. Sol\'e, \emph{Proc. R. Soc. Lond. B} {\bf 268}, 2261 (2001).

\bibitem[S2]{caldeira} S.M.G. Caldeira, T.C. Petit Lob\~ao, R.F.S. Andrade, A. Neme, and J.G.V. Miranda, \emph{Eur. Phys. J. B} {\bf 49}, 523 (2006).

\bibitem[S3]{motter} A. Motter, A.P.S. De Moura, Y.C. Lai, and P. Dasgupta, \emph{Phys. Rev. E}, {\bf 65}, 065102 (2002).

\bibitem[S4]{trifonov86} V. Brendel, J.S. Beckmann, and E.N. Trifonov, \emph{Journal of Biomolecular Structure \& Dynamics} \textbf{4}, 011 (1986).

\bibitem[S5]{caselle} M. Caselle, F. Di Cunto, and P. Provero, \emph{BMC Bioinformatics} {\bf 3}, 7 (2002); D. Cor\`a, F. Di Cunto, P. Provero, L. Silengo, and M. Caselle, \emph{BMC Bioinformatics} {\bf 5}, 57 (2004). 

\bibitem[S6]{nicodeme} P. Nicod\`eme, T. Doerks, and M. Vingron, \emph{Bioinformatics} {\bf 18}: S161, Suppl.2 (2002).

\bibitem[S7]{giansanti} A. Giansanti, M. Bocchieri, V. Rosato, and S. Musumeci, \emph{Parasitol. Res.} {\bf 101}, 639 (2007).

\bibitem[S8]{CCDS} Data downloaded from the \emph{Consensus Coding Sequence database} (CCDS), http://www.ncbi.nlm.nih.gov/CCDS/, version Hs35.1. 

\bibitem[S9]{average} With the notation $\left\langle p \left( x \right) \right\rangle $, we denote the average of $p(x)$ over all the possible configurations of $x$ and with $\sigma$ the standard deviation of the distribution.

\bibitem[S10]{vandongen} A. J. Enright, S. Van Dongen, and C. A. Ouzounis, \emph{Nucleic Acids Research} {\bf 30}:1575 (2002).

\bibitem[S11]{prosite} http://www.expasy.ch/prosite

\bibitem[S12]{article_preparation} R. Sinatra, D. Condorelli, A. Giansanti, V. Latora, V. Rosato, \emph{Analysis of proteomes by means of $k$-motif networks}, in preparation.

\bibitem[S13]{leet} http://en.wikipedia.org/wiki/Leet

\bibitem[S14]{twitter} www.twitter.com

\bibitem[S15]{twitter_cascades} K. Lerman and R. Ghosh, in Proc. of ICWSM (2010).

\bibitem[S16]{flicker_cascades} M. Cha, A. Mislove, B. Adams, K. P. Gummadi, in Proc. of WOSN 08 - USA (2008).

\bibitem[S17]{API} http://apiwiki.twitter.com/Streaming-API-Documentation

\bibitem[S18]{cohen} http://en.wikipedia.org/wiki/Benjamin\_Cohen\_\%28journalist\%29

\bibitem[S19]{screengrab} http://twitpic.com/1jge7b

\bibitem[S20]{duffy} http://www.guardian.co.uk/politics/2010/jul/26/gillian-duffy-backs-david-miliband

\bibitem[S21]{tactical} http://wiki.electorama.com/wiki/Tactical\_voting

\bibitem[S22]{chirikov} B.V. Chirikov, Phys. Rep. {\bf 52}:263 (1979).

\bibitem[S23]{aleksev} V.M. Aleksev and M.V. Yakobson, \emph{Phys. Rep.} {\bf 75}:287 (1981).


\end{thebibliography}
\end{document}